\documentclass[a4paper,pdftex,12pt]{article}

\usepackage{ifpdf}
\usepackage{geometry}
\usepackage[utf8]{inputenc} 
\usepackage[T1]{fontenc}    
\usepackage{hyperref}       
\usepackage{url}            
\usepackage{booktabs}       
\usepackage{amsfonts}       
\usepackage{nicefrac}       
\usepackage{microtype}      
\usepackage[pdftex]{graphicx}
\usepackage{amsmath,thmtools}
\usepackage{subfigure}
\usepackage{amsmath}
\usepackage{bm}
\usepackage{multirow}
\usepackage{float}
\usepackage[misc]{ifsym} 
\usepackage{color}
\usepackage[numbers]{natbib}
\usepackage{inputenc}
\usepackage{silence }
\usepackage{cleveref}
\usepackage{lineno}

\title{A novel convolutional neural network model to remove muscle artifacts from EEG}

\author{Haoming Zhang$^{1,\dagger}$, Chen Wei$^{1,\dagger}$, Mingqi Zhao$^{2}$, \\Haiyan Wu$^{3}$, Quanying Liu$^{1,\ast}$}

\begin{document}
%
\maketitle

\vspace{-0.7cm}
\begin{centering}
$^1$ Department of Biomedical Engineering, Southern University of Science and Technology, Shenzhen 518055, P.R. China\\
$^2$ Movement Control and Neuroplasticity Research Group, KU Leuven, Leuven 3001, Belgium \\
$^3$ Center for Cognitive and Brain Sciences, University of Macau, Taipa, Macau, China\\

\vspace{0.5cm}
$^\dagger$  Equal contribution\\
$^\ast$ Corresponding author \texttt{liuqy@sustech.edu.cn} to Q.L.\\
\vspace{0.5cm}

\end{centering}
\section*{Abstract}

The recorded electroencephalography (EEG) signals are usually contaminated by many artifacts. In recent years, deep learning models have been used for denoising of electroencephalography (EEG) data and provided comparable performance with that of traditional techniques. However, the performance of the existing networks in electromyograph (EMG) artifact removal was limited and suffered from the over-fitting problem. Here we introduce a novel convolutional neural network (CNN) with gradually ascending feature dimensions and downsampling in time series for removing muscle artifacts in EEG data. Compared with other types of convolutional networks, this model largely eliminates the over-fitting and significantly outperforms four benchmark networks in EEGdenoiseNet. Our study suggested that the deep network architecture 
might help avoid overfitting and better remove EMG artifacts in EEG.\\

Keywords: Convolutional neural network, Electroencephalography, Muscle artifact removal, EEG denoising

\section{Introduction}
\label{sec:intro}

Electroencephalography (EEG) measures the electrical potential over the scalp, which is widely used in both cognitive neuroscience and brain-computer interface \cite{cai2020feature-level,oberman2005eeg,wolpaw1991an,herrmann2005human,liu2017detecting,zhao2019hand}. The raw electroencephalogram (EEG) data are often contaminated by various noise and physiological artifacts, such as ocular artefacts
\cite{croft2000removal}, myogenic artefacts \cite{muthukumaraswamy2013high-frequency, piontonachini2019iclabel:}, and cardiac artefacts \cite{marino2018adaptive}. Therefore, artifacts removal is an essential step for the application of EEG technique.


Ocular and cardiac artefacts have simple features and therefore easy to be dealt with. On the contrary, myogenic artifacts have numerous sources and their frequency spectrum largely overlaps with the frequencies of interest in EEG signals. As a result, muscle artifacts induced by muscle contractions are particularly difficult to remove using traditional methods, such as adaptive filtering. Thus, data-driven approaches might be potential to extract the features of the neural signals from the muscle artifacts contaminated EEG signals and reconstruct the pure EEG signals. 

\section{Related work}
\label{sec:related_work}
Although deep learning (DL) has been widely used in computer vision and natural language process (NLP), the DL methods for EEG denoising is an emerging field. To the best of our knowledge, we only found four DL-based studies in EEG denoising \cite{yang2018automatic, sun2020novel, Hanrahan2019NoiseRI, eegdenoisenet}. They offered comparable performance with that of traditional denoising techniques, especially for EOG artifacts. Previous studies have reported the application of 
a 5-layer neural network in removing ocular artifacts~\cite{yang2018automatic}, a convolutional autoencoder~\cite{Hanrahan2019NoiseRI} and a novel end-to-end 1D-ResCNN model to remove multiple types of artifacts~\cite{sun2020novel}. More recently, a benchmark dataset for deep learning solutions of EEG denoising, EEGdenoiseNet , has been proposed. This benchmark dataset consists of a large number of clean EEG, and pure EOG and EMG signal epochs for training and testing DL models, as well as benchmarking networks, including a fully-connected network (FcNN), a simple convolution network, a complex convolution network and a recurrent neural network (RNN). These networks worked well on ocular artifacts, but not on myogenic artifacts. Particularly, the two convolutional networks suffered severe over-fitting problem, which limits the generalizability of the networks in test data. Thus, much work need to be done for myogenic artifact removal.

In this study, we propose a novel convolutional neural network (Novel CNN) to improve denoising performance for EMG artifacts and to enhance the generalization ability. Our results show that the Novel CNN could effectively reconstruct the denoised EEG signals through the multiple convolutional layers and a single fully connected layer. We believe our study can pave the path of convolutional networks in the area of EMG noise reduction.
\section{Method}
\label{sec:method}

\subsection{Generation of training, validation and test dataset}
We used data from EEGdenoiseNet to generate pairs of pure and noisy EEG signals for training and testing the proposed neural network. Specifically, 4514 EEG epochs and 5598 EMG epochs were used to simulate noisy EEG with myogenic artifacts. We randomly reused some of the data to increase the number of EEG epochs to 5598 and obtained 5598 pairs of EEG and myogenic artifact epochs. We randomly divided 5598 pairs of data into 10 parts, 8 parts for training set (4478 pairs), 1 part for validation set (560 pairs) and 1 part for test set (560 pairs). For the training set, we randomly combined 4478 pairs of EEG and myogenic artifact data ten times by linearly mixing the EEG epochs with EMG epochs according to eq.\ref{formula: EEG mixing}, using the signal to noise ratios (SNRs, see eq.\ref{formula: SNR}) sampled from a uniform distribution from -7dB to 2dB. In the formulas, $y$ denotes the mixed signal of EEG and myogenic, $x$ denotes the clean EEG signal, $n$ denotes myogenic, and $\lambda$ denotes the relative contribution of EMG artifact.For the validation set and test set, we combined 560 pairs of epochs with the same levels of SNR as training set, and expanded the validation set and testing set to 5600 epochs.

\begin{equation}
y = x +\lambda \cdot n\label{formula: EEG mixing}
\end{equation}

\begin{equation}
SNR =10\log {\frac{RMS(x)}{RMS(\lambda \cdot n)}}\label{formula: SNR}
\end{equation}

\subsection{Network structure}
\begin{figure}[ht]
\centering
\includegraphics[width=0.9\textwidth]{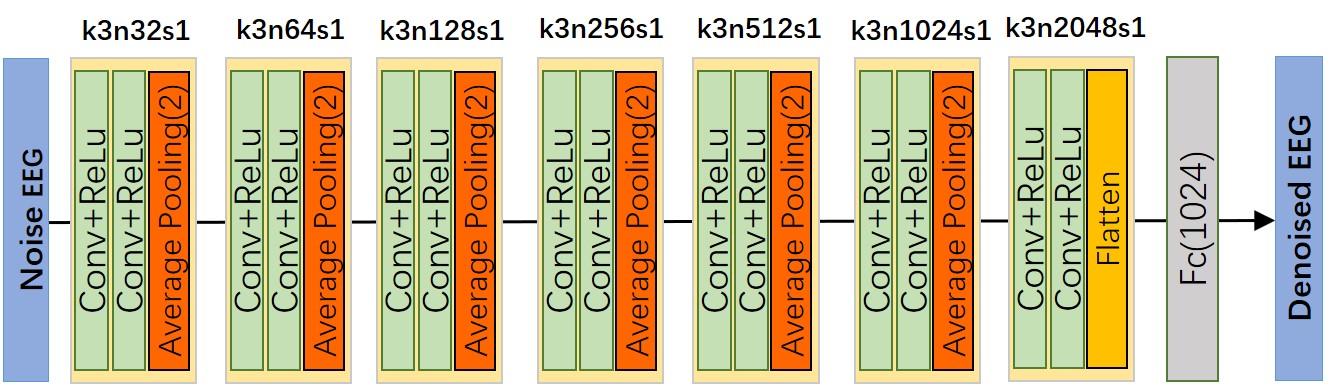}
\caption{The structures of the Novel CNN.}
\label{fig:novel_CNN1}
\end{figure}

\begin{figure}[ht]
\centering
\includegraphics[width=0.9\textwidth]{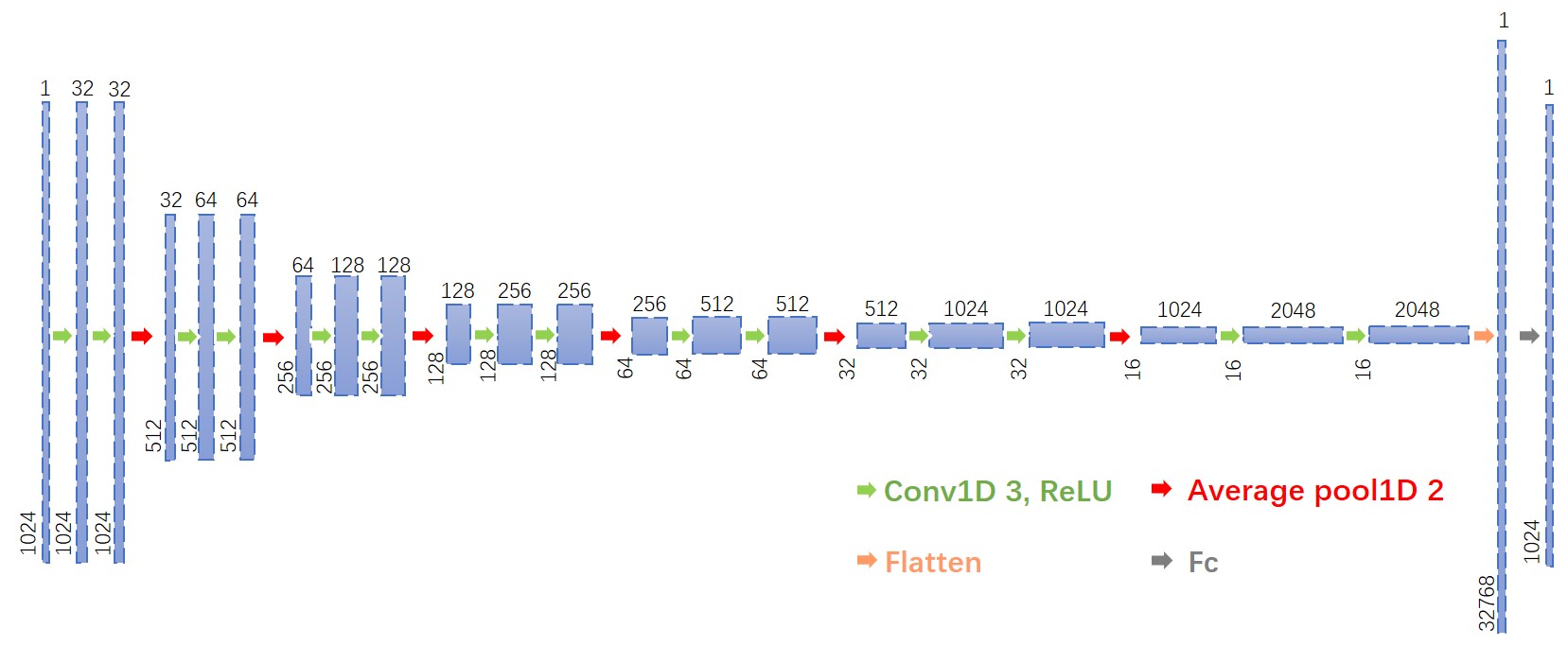}
\caption{The size of the EEG signal in each layer. }
\label{fig:novel_CNN2}
\end{figure}

The proposed Novel CNN for myogenic artifact reduction (see Fig. \ref{fig:novel_CNN1})  contains seven similar blocks. In each of the first six blocks, two 1D-convolution layers with small 1*3 kernels, 1 stride, and a ReLU activation function are followed by a 1D-Average pooling layer with pool size equal to two. In the seventh block, two 1D-convolution layers are followed by a flatten layer. The last block was followed by a dense layer with 1024 outputs, the same dimension as the input. In order to gradually extract features and increase feature dimension in the hierarchical network, the number of feature maps of the convolutional layer in each block follows an exponential function from 32 to 2048 (32 64 128 256 512 1024 2048). The network gradually reduce the EEG signal sampling rate by the 1D-Average pooling layer. We  visualized the size of the EEG signal as it passes through each layer (see Fig. \ref{fig:novel_CNN2}). 

We compared the performance of Novel CNN with the 4 benchmark networks in EEGdenoiseNet, \textit{i.e.} a fcNN, simple CNN, a RNN and a complex CNN. The structure details of the 4 benchmark networks can be found in Fig.5 of EEGdenoiseNet\cite{eegdenoisenet}. 

\subsection{Learning process}
To enable the flexibility of the neural network for a large range of EEG amplitudes, we adopted the idea from the EEGdenoiseNet and normalized the input noisy EEG signal, $y$, and the ground-truth EEG signal, $x$, by dividing the standard deviation of noisy EEG signal according to eq. \ref{formula:normalization}:

\begin{align}
\label{formula:normalization}
\hat{x} = \frac{x}{\sigma_{y}},  &  &  \hat{y} = \frac{y}{\sigma_{y}}
\end{align}
where $\sigma_{y}$ is the standard deviation of noisy signal $y$, which will multiply with corresponding denoised output EEG signals to restore the original magnitude of EEG. $\hat{y}$ will input to neural networks.

The goal of a noise reduction network is to formulate a nonlinear function $f$ that projects normalized noisy EEG $\hat{y}$ to the cleaned EEG $\tilde{x}$ (see eq. \ref{formula: map function}):
\begin{equation}
\tilde{x} = f(\hat{y},\theta)
\label{formula: map function}
\end{equation}
where $\hat{y}\in \mathbb{R}^{1\times T}$ denotes the normalized noisy signal as the input, $\tilde{x}\in \mathbb{R}^{1\times T}$ denotes the cleaned EEG signal as the output, and $\theta$ is the parameters to be learned. Considering $\hat{x}$ as a sample from the clean EEG signal distribution $P(\hat{x})$ and $\hat{y}$ as a sample from the noisy EEG signal distribution $P(\hat{y})$, the noise reduction network can also be described as a function $f$ that maps samples from $P(\hat{y})$ to a distribution $P(\tilde{x})$, with the difference between $P(\tilde{x})$ and $P(\hat{x})$ is minimized. The denoising network could move the noisy EEG signal distribution $P(\hat{y})$ to the clean EEG signal distribution $P(\hat{x})$.

We used the mean squared error (MSE) as loss function $L_{MSE}(f)$ (see eq. \ref{formula: loss}). The learning process was implemented with gradient descent to minimize the difference between noisy and ground.

\begin{equation}
L_{MSE}=\frac1{N}\sum_{i=1}^N\Big|\Big|\tilde{x}_i-\hat{x}_i\Big|\Big|^2_2\label{formula: loss}
\end{equation}
where $N$ denotes the number of samples of an epoch, and $\hat{x}_i$ denotes the $i^{th}$ sample of the normalized ground truth epoch $x$, $\theta$ is the parameters of network, and $\tilde{x}_i$ is the output of the Novel CNN as the cleaned EEG.

We trained the Novel CNN with 50 epochs, and the CNN models were optimized by the RMSprop optimizer, with the hyperparameters set as $\alpha = 5e^{-5}$, $\beta=0.9$. To increase the statistical power of our results, the Novel CNN, as well as other benchmark networks, were trained, validated and tested independently for 10 times with randomly generated datasets, in the same manner as our previous study \cite{eegdenoisenet}. 

All networks were implemented in Python 3.7 with Tensorflow 2.2 library, running on a computer with two NVIDIA Tesla V100 GPUs. The codes for the Novel CNN algorithms is publicly available online at \url{https://github.com/ncclabsustech/EEGdenoiseNet/tree/master/code/Novel_CNN}.

\subsection{Performance Evaluation}
To compare the performance of Novel CNN with the benchmark networks in EEGdenoiseNet, we used the same performance evaluation. 
First, network convergence was employed to provide  information about the learning, testing and diagnose of the networks. The convergence curve of both training and test processes was visualized by calculating the average loss function value (see eq.\ref{formula: loss}) with respect to the number of epochs.

To examine the performance of the networks, we applied three objective measures \cite{2017Independent} on the denoised data, including Relative Root Mean Square Error (RRMSE) in the temporal domain ($RRMSE_{t}$, see eq. \ref{formula: RRMSEt}), RRMSE in the spectral domain ($RRMSE_{f}$, see eq. \ref{formula: RRMSEf}) and the average correlation coefficient ($CC$ see eq. \ref{formula: ACC}).
\begin{equation}
RRMSE_{t} =\frac{RMS(f( y ) -x)}{RMS(x)}\label{formula: RRMSEt}
\end{equation}

\begin{equation}
RRMSE_{f} =\frac{RMS(PSD(f(y))-PSD(x))}{RMS(PSD(x))}\label{formula: RRMSEf}
\end{equation}
where the function $PSD( )$ denotes to the power spectral density of an input signal.

\begin{equation}
CC =\frac{Cov(f(y),x)}{\sqrt{Var(f(y))Var(x)}}\label{formula: ACC}
\end{equation}

\section{Results}
\label{sec:results}

\begin{figure}[ht]
\centering

\includegraphics[width=0.9\textwidth]{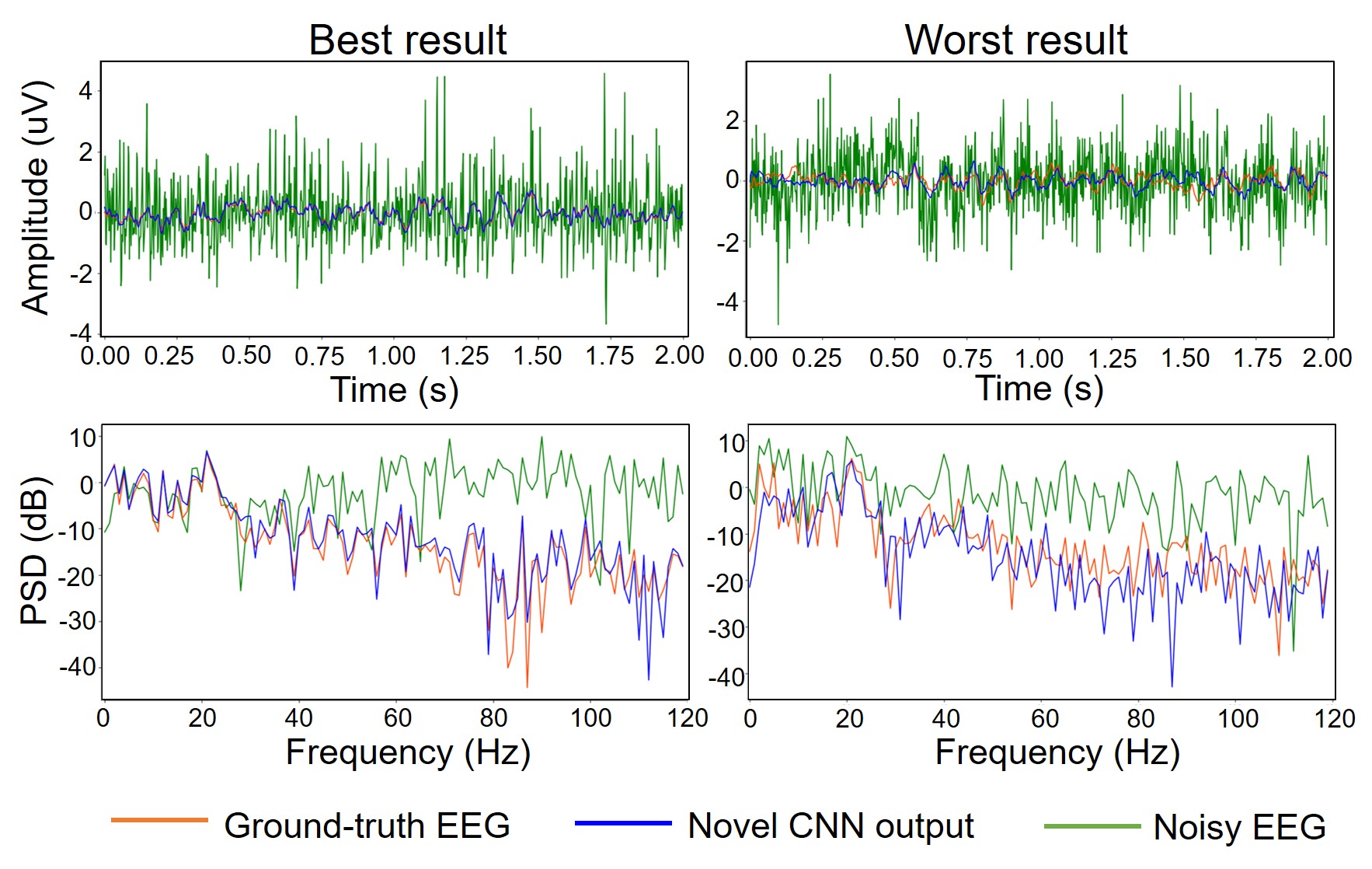}
\caption{Some examplary epochs showing the performance in temporal (upper) and spectral (bottom) domains for Novel CNN. (left) The examples with the best denoising performance; (right) the examples with the worst denoising performance. The green, red and blue line represent the noisy EEG, the ground-true EEG and the cleaned EEG  respectively.}
\label{fig:EMG_novelCNN_examples}
\end{figure}

We first demonstrated two examples (one best and one worst) of the myogenic artifact removal results of our Novel CNN in the test set (see Fig. \ref{fig:EMG_novelCNN_examples}). The SNR of these  input noisy EEG signals is -6dB. We observed that the high frequency artefacts were largely attenuated in both cases. The output of the best case generally looks close to ground-truth EEG, whereas the output of the worst case showed poor correlation with ground-truth EEG.

\begin{figure}[ht]
\centering
\includegraphics[width=0.9\textwidth]{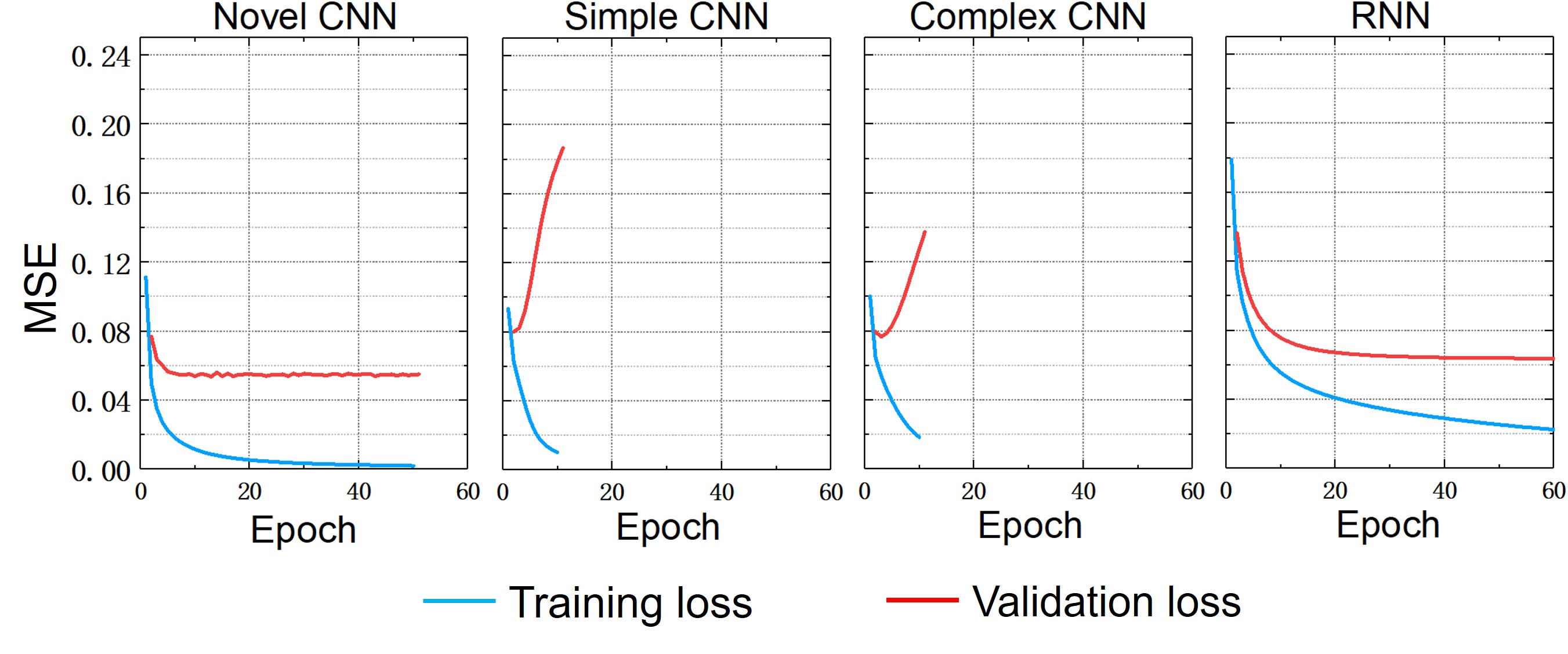}
\caption{Mean MSE loss of the training set and test set versus the number of epochs in myogenic artifact removal.}
\label{fig:EMG_loss}
\end{figure}

We then dispay the loss versus the number of epochs for the four networks in myogenic artifact removal (see Fig. \ref{fig:EMG_loss}), including Novel CNN, and three benchmark networks in EEGdenoiseNet (Simple CNN, Complex CNN and RNN). In general, Novel CNN and RNN showed decreasing trend in both training loss and test loss, whereas Simple CNN and Complex CNN presented a serious over-fitting and reached their minimum values after one epoch and two epochs respectively. Compared to the two kinds of convolutional networks in benchmark, overfitting did not occur during the training process of Novel CNN, and it's lowest value of test loss is significantly lower. Furthermore, the test loss and train loss of Novel CNN are all lower than RNN, which indicate Novel CNN has better performance than RNN in both training and test set.

\begin{figure}[ht]
\centering
\includegraphics[width=0.8\textwidth]{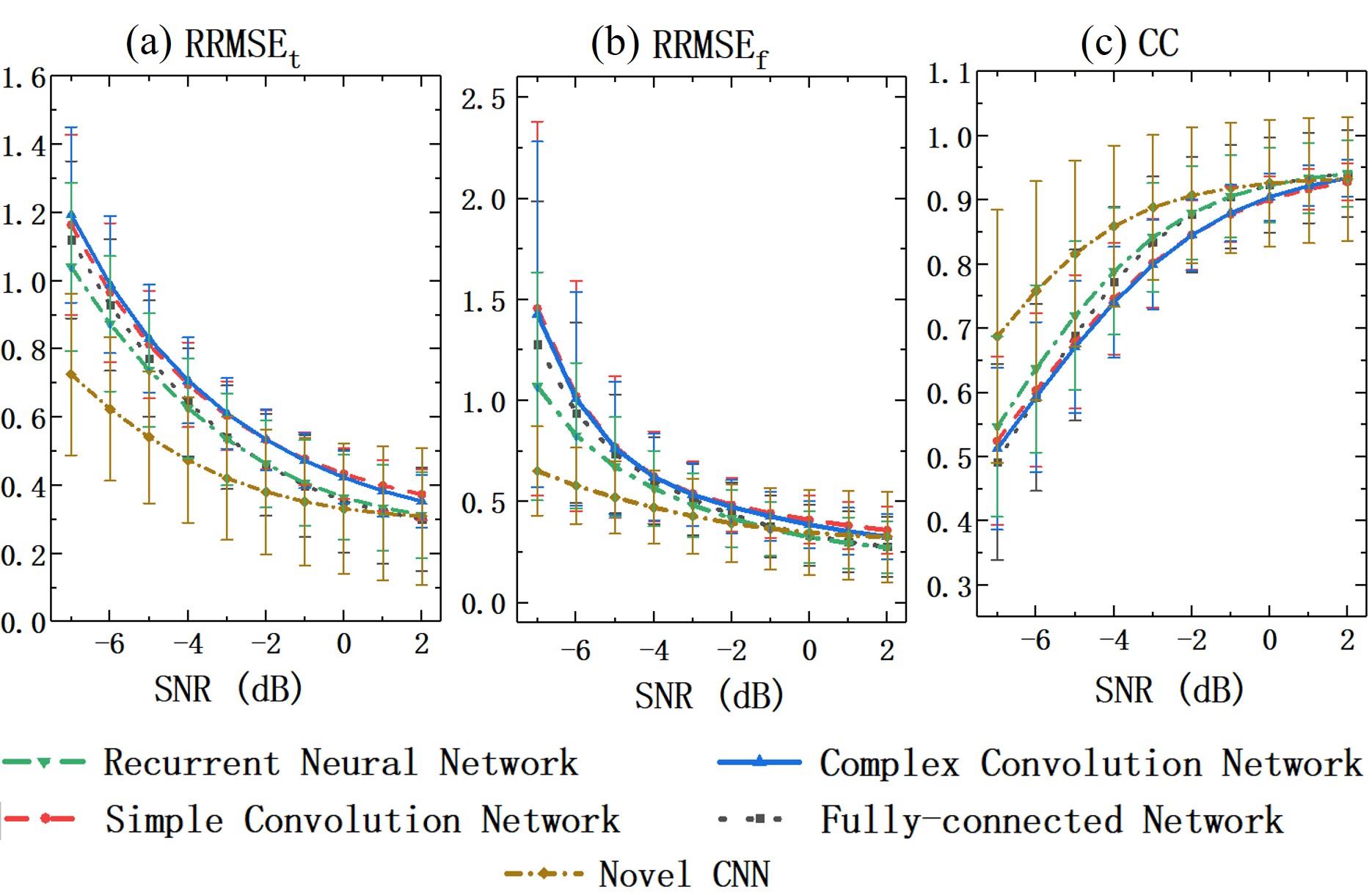}
\caption{Performances of five networks in myogenic artifact removal with different SNRs. (a) $RRMSE_{t}$. (b) $RRMSE_{f}$. (c) CC.}
\label{fig:EMG_evaluation}
\end{figure}

Fig. \ref{fig:EMG_evaluation} displays the three evaluation parameters to quantitative evaluate myogenic artifact removal capability of Novel CNN and four benchmark networks in different noise levels. Among the five networks, Novel CNN shows the lowest $RRMSE_{t}$, $RRMSE_{f}$ and the highest $CC$ value in high noise level and middle noise level. But in low noise level, FcNN and RNN show lower $RRMSE_{t}$ (at SNR = 2dB), $RRMSE_{f}$ (at SNR  \textgreater -1dB) and higher $CC$ (at SNR \textgreater 1dB). The values of the FCNN and RNN show little difference.

We further summarized the quantitative benchmarks by averaging the values over all SNRs (see Table \ref{table: performance_EMG}). As shown in Tables 1, the Novel CNN had the lowest average $RRMSE_{t}$, $RRMSE_{f}$ and highest average $CC$. From the values of the three quantitative indicators, we could clearly summarized that the performance of Novel CNN significantly better than the four benchmark networks.
\begin{table}
  \caption{Average performances of all SNRs in myogenic artifact removal.}
  \label{table: performance_EMG}
  \centering
  \begin{tabular}{llll}
    \toprule
    \cmidrule(r){1-2}
    Model   & $RRMSE_{t}$  & $RRMSE_{f}$    &  CC \\
    \midrule
    FCNN        & $    0.585 $ & $    0.580 $ & $    0.796 $ \\
    \small{Simple CNN}  & $    0.646 $ & $    0.649 $ & $    0.783 $\\
    \small{Complex CNN} & $    0.650 $ & $    0.633 $ & $  0.780 $\\
    RNN         & $0.570$ & $0.530$ & $0.812$\\
    $\textbf{Novel  CNN}$ & $\bm{0.448}$ & $\bm{0.442}$ & $\bm{0.863}$\\
    \bottomrule
  \end{tabular}
\end{table}


\section{Discussion}
\label{sec:discussion}




Convolutional network are often applied in the field of computer version and image processing. It also performs well in other fields. For example, CNN has been used for text classification in NLP \cite{textcnn, 2015Character}, and their performance is comparable to RNN based network. Similar to nature language, EEG and EMG are also time series data, hence we could assume that CNN also has great potential in the field of EEG noise reduction; but the fact is that in our previous study \cite{eegdenoisenet}, both simple CNN and complex CNN showed severe generalization issue in the test set. The resolving of this problem might be the key to improve the performance of CNN in similar application scenarios. 

To propose a novel CNN that addresses the generalization issue of the CNNs \cite{eegdenoisenet, sun2020novel}, we first tried to find clues that related to this problem. By comparing the structure of the two CNN networks (simple \cite{eegdenoisenet} and complex \cite{sun2020novel}) with ResNet  \cite{he2016deep}, VGGNet \cite{2014Very} and U-net \cite{unet} that do not suffer from generalization problem, we suspected that the number of feature maps might be the reason that limits the generalization performance of the two CNNs, since ResNet, VGGNet and U-net are several times larger in feature map size than the two CNNs. Inspired by ResNet, VGGNet and U-net, we therefore increased the number of feature maps with the increase of network depth, and downsampled the signal in temporal field  to reduce the number of parameters using Averagepooling to prevent overfitting  \cite{Cao2015APT}. Remarkably, our Novel CNN overcame the generalization problem, and also showed better performance in general than the four benchmark networks in our previous study \cite{eegdenoisenet}. This is not surprising, since the separation of the features of brain activity from myogenic activity is particular difficult, 
while the parameter of each layer of the Novel CNN was elaborately designed instead of simply stacked the convolutional layers with the same parameter together, which improves such separation
and overcome the generalization problem and improved the performance of the network in myogenic artifact removal.

Several limitations of our studies should be noted. First, the dataset we used to train the Novel CNN, EEGdenoisenet, contains only thousands of EMG epochs. These might not be sufficient to learn the complex features of MEG using neural networks. Second, inherited from the benchmark networks in EEGdenoisenet, Novel CNN only work on the 2s-long EEG epochs. A better design for longer EEG signals is needed in practical applications. One possible approach could be an additional network block to the existing networks, which allows the learning of the hidden relationship between continuous epochs for continuous muscle artifact reduction. Third, our Novel CNN remarkably outperforms RNN with high-level noise, but it works worse than RNN at low noise levels. In the future, we will continue to improve the CNN to better differentiate the features of neural signals from myogenic electrical artifacts, and obtain the denoised EEG.

\vfill \pagebreak


\bibliographystyle{unsrt}
\bibliography{references}

\end{document}